\documentclass[12pt]{article}

\def\figurename{Figure}
\usepackage{amsmath}
\usepackage{rotating}
\usepackage{graphicx}
\usepackage{lscape}
\usepackage[utf8]{inputenc}
\usepackage{amsfonts}
\usepackage{amssymb}
\usepackage{amsthm}
\usepackage[T1]{fontenc}
\usepackage[spanish]{babel}
\usepackage[active]{srcltx}
\usepackage{enumerate}
\usepackage{graphicx}
\usepackage{epsfig}
\usepackage{pb-diagram}
\usepackage{tensor}
\usepackage{bm}

\begin{document}
\newcommand{\nl}{{\cal N}}
\newcommand{\cl}{\Phi}
\def\binom#1#2{{#1\choose #2}}
\newcommand{\sas}{ $\hat{S}^{2}$ and $\hat{S}_{z}$ }
\newcommand{\hu}{{\cal A}}
\newcommand{\dm}{{\cal D}}
\newcommand{\idn}{{\cal I}}
\newcommand{\lo}{\Lambda\Omega}

\renewcommand{\figurename}{\textbf{\textit{Figure}}}

%%%%% EQUATIONS %%%%%%%%%%%%%%%%%%%%%%%%%%%%%%%%%%%

%\numberwithin{equation}{section}

%%%%% THEOREMS %%%%%%%%%%%%%%%%%%%%%%%%%%%%%%%%%%%%%

\newtheorem{corollary}{Corollary}%[section]
\newtheorem{lemma}{Lemma}%[section]
\newtheorem{proposition}{Proposition}%[section]
\newtheorem{theorem}{Theorem}%[section]

%%%%% demS %%%%%%%%%%%%%%%%%%%%%%%%%%%%%%%%%%%%%%%%

\newenvironment{dem}[1][Proof]{\begin{proof}[{\bf #1}]}{\end{proof}}

%%%%% DEFINITIONS %%%%%%%%%%%%%%%%%%%%%%%%%%%%%%%%%%%%

\theoremstyle{definition}

\newtheorem{axiom}{Axioma}[section]
\newtheorem{definition}{Definición}[section]
\newtheorem{example}{Ejemplo}[section]
\newtheorem{remark}{Observación}[section]
\newtheorem{exercise}{Ejercicio}[section]
{\swapnumbers\newtheorem{exercisestar}[exercise]{(*)}}

%%%%% NUMERICAL SETS %%%%%%%%%%%%%%%%%%%%%%%%%%%%%%%%

\newcommand{\C}{\ensuremath{\mathbb{C}}}
\newcommand{\N}{\ensuremath{\mathbb{N}}}
\newcommand{\Q}{\ensuremath{\mathbb{Q}}}
\newcommand{\R}{\ensuremath{\mathbb{R}}}
\newcommand{\T}{\ensuremath{\mathbb{T}}}
\newcommand{\Z}{\ensuremath{\mathbb{Z}}}

%%%%% FUNCTIONS %%%%%%%%%%%%%%%%%%%%%%%%%%%%%%%%%%%%%

\newcommand{\abs}[1]{\left\vert #1\right\vert}
\newcommand{\bra}[1]{\left\langle #1\right\vert}
\renewcommand{\dim}[1]{\mathrm{dim}\left( #1\right)}
\newcommand{\set}[1]{\left\{ #1\right\}}
\renewcommand{\ker}[1]{\mathrm{Ker}\pa{#1}}
\newcommand{\ket}[1]{\left\vert #1\right\rangle}
\newcommand{\norm}[1]{\left\|#1\right\|}
\newcommand{\pa}[1]{\left(#1\right)}
\newcommand{\pro}[2]{\left\langle#1|#2\right\rangle}
\newcommand{\proo}[3]{\left\langle#1\left|#2\right|#3\right\rangle}
\newcommand{\ran}[1]{\mathrm{Ran}\pa{#1}}
\newcommand{\tr}[1]{\mathop{\mathrm{Tr}}\pa{#1}}

\newcommand{\Rn}[1][\mathcal{N}]{\mathbb{R}^{#1}}
\newcommand{\Cn}[1][\mathcal{N}]{\C^{#1}}
\newcommand{\Sph}[1][\mathcal{N}]{\mathrm{S}^{#1}}

\newcommand{\codim}[1]{\mathrm{codim}\left( #1\right)}
\newcommand{\diag}{\mathop{\mathrm{diag}}}
\newcommand{\diam}{\mathop{\mathrm{diam}}}
\newcommand{\id}{\mathop{\mathrm{id}}}
\newcommand{\inte}{\mathop{\mathrm{int}}}
\newcommand{\inc}{\mathop{\mathrm{\iota}}}
\newcommand{\sop}{\mathop{\mathrm{sop}}}

\newcommand{\GL}[2][R]{\mathrm{GL}\pa{\mathbb{#1},#2}}
\newcommand{\GLp}[2][R]{\mathrm{GL}_{+}\pa{\mathbb{#1},#2}}
\newcommand{\SL}[2][R]{\mathrm{SL}\pa{\mathbb{#1},#2}}
\newcommand{\Or}[1][\mathcal{N}]{\mathrm{O}\pa{#1}}
\newcommand{\SO}[1][\mathcal{N}]{\mathrm{SO}\pa{#1}}
\newcommand{\Un}[1][\mathcal{N}]{\mathrm{U}\pa{#1}}
\newcommand{\SU}[1][\mathcal{N}]{\mathrm{SU}\pa{#1}}
\newcommand{\Up}[2][R]{\mathrm{Up}\pa{\mathbb{#1},#2}}
\newcommand{\Upp}[2][R]{\mathrm{Up}_{+}\pa{\mathbb{#1},#2}}
\newcommand{\her}[1]{\mathrm{her}\pa{#1}}
\newcommand{\hil}{\mathsf{H}}

\renewcommand{\Re}[1]{\mathrm{Re}\left( #1\right)}
\renewcommand{\Im}[1]{\mathrm{Im}\left( #1\right)}

 %\DeclareMathOperator{\sgn}{sgn} \DeclareMathOperator{\Ln}{Ln}
%\DeclareMathOperator{\card}{\#}

%%%%%%%%%%%%%%%%%%%%%%%%%%%%%%%%%%%%%%%%%%%%%%%%%%%%%%%%%%%%%%%%%%%%%%%%%%%

%\vskip -10mm

% \vskip -10mm

{\Large \noindent {\bf Electron distributions of molecular domains: canonical ensemble,
and charge transfer electronegativity relationship}}

\vskip 5mm

\noindent {\bf Roberto C. Bochicchio}$^{*,1,2}$

 \vskip 2mm

{\small

\noindent $^{1}$ {\it Universidad de Buenos Aires, Facultad de
Ciencias Exactas y Naturales, Departamento de F\'{\i}sica, Ciudad
Universitaria, 1428, Buenos Aires, Argentina}

\noindent $^{2}$ CONICET - {\it Universidad de Buenos Aires,
Instituto de F\'{\i}sica de Buenos Aires (IFIBA) Ciudad
Universitaria, 1428, Buenos Aires, Argentina}

}

\vskip 10mm

%\normalsize
% \noindent {\bf Abstract \hspace{0.25cm}}

\noindent The principle of maximum entropy (MaxEnt) applies to the
canonical ensemble related to the number of particles, known as the
$\mathcal{N}$-ensemble. This concept pertains to physical domains (or basins) 
that are treated as open systems capable of transferring charge through the 
exchange of electrons. In this context, fractional occupation numbers of electrons 
indicate a net charge, represented as $\nu$. This principle outlines the convex expansion of the 
density matrix (DM), based on three distinct electronic states: the neutral state and 
two ionic edge states, each with a charge limit of $\pm q$. The coefficients of expansion 
and the charge transference fraction $\nu$, are crucial for understanding electron distribution. We 
express the quantities discussed as functions of the chemical potential derived from the statistical 
ensemble. Our analysis focuses on the donor and acceptor characteristics of different domains 
in relation to these parameters. The physical compatibility of the current equilibrium states of the 
system is discussed within this statistical framework and the electronegativity of the domains is 
inferred from the physical behavior of the associated populations.

\vskip 40mm

{\small \noindent
\rule{60mm}{0.4mm} \\
\noindent $^{*}$ rboc@df.uba.ar}

\newpage
\noindent {\large {\bf I. INTRODUCTION}}

\vskip 3mm

Molecular systems (S) consist of interacting charged particles,
including electrons and nuclei, and are primarily influenced by
Coulomb interactions. This type of interaction in these systems has
an important experimental feature: their energies and, consequently,
their electronic densities are convex functions of the number of
particles when viewed as a closed system with a fixed integer number
of particles. $^{1-4}$

The convexity property of these systems allows their ground state
Density Matrix $D$ (DM) to be rigorously spanned by a two-state
model within a two-dimensional Fock space. This representation
involves the quantum state of the neutral species {\bf X}$^{0}$ and
only one of the ionic species, {\bf X}$^{+}$ or {\bf X}$^{-}$,
depending on whether the population is greater or lower than that of
the neutral species, respectively. This has been conjectured in Ref. $^{1}$  
and rigorously derived in terms of DMs in Ref. $^{4}$. Energy 
discontinuities arise within this framework at integer number of particles. $^{2}$  
Furthemore, under a perturbation interaction with an environment (R), such as a reservoir or a 
solvent field, the energy may change only slightly, then despite this perturbation, the
energy still exhibits a convex relationship with the number of
particles, which means it continues to satisfy the mentioned convexity
property and then it avoids the energy and density derivative discontinuities 
when determining chemical descriptors as chemical potential and hardness.$^{5,6}$

The molecular structure considered through the rigorous Quantum
Theory of Atoms in Molecules (QTAIM), $^{7,8}$ is composed of a set of
non-overlapping physical regions or basins $\Omega$, called domains. These
physical domains within the molecular system can be represented by an atom, a functional group, 
or simply a moiety that interacts with the other domains in the system. They are really subsystems 
interacting between them exchanging charge (electrons), so these domains can accommodate a fractional 
number of electrons, i.e., electron populations. As a result, the number of particles in the domain, 
denoted as $\nl$, must be treated as a continuous variable. Thus, this domain
should be regarded as an open quantum system. $^{5,9,10}$ Unlike systems that have an
integer number of particles $N$, this type of systems lacks the
previously mentioned convexity. In this scenario, since the convexity hypothesis does not hold, this
open system requires a significant convex expansion of $D$. $^{11}$ This is the key distinction between
a complete molecular system {\bf X}$^{0}$ and/or its ionic forms {\bf X}$^{\pm}$ and the physical domains 
$\Omega$ within these systems.

This expansion must include several pure states,
specifically more than two, for a precise description. Each of these states are 
characterized by its integer particle number $N$, as previously
demonstrated. $^{12}$ This refers to a description of
the statistical distribution of electrons within molecules. By
employing an ensemble statistical distribution formalism for systems of a few 
number of particles, we were able to create a more comprehensive
quantum state for the new non-convex scenario while only considering the
electron population in the domain $\nl$. $^{1,2,4}$ To pursue this goal,
we employed an algebraic model consisting of three states: one
neutral state (denoted as $N$ electrons) and two ionic edge states.
These ionic states are characterized by their maximum acceptor and
donor capacities, represented as $N \pm q$, where $q \in
\mathbb{Z}^+$ indicates the maximum number of electrons that can be
accepted or donated by the neutral species, respectively. This
approach enabled us to systematically derive all quantum states of
the system, including both equilibrium and non-equilibrium states. 
$^{12}${BochMaulen2023}

The current report continues previous research by implementing
maximum entropy equilibrium states $D$ for a domain, aiming to
determine the expansion coefficients with the only constraint being
the number of particles $\nl$. This approach enables us to express
the charge transference fraction $\nu$, which is considered the net
charge in the domain, as a function of the statistical coefficients
defining the quantum state of the domain and a key physical
parameter known as the chemical potential. Additionally, it
highlights the relationship between these elements, illustrating the
domain's donor or acceptor characteristics. This is the primary
objective of the present report.

It is essential to note that the current approach does not take into
account a stabilization process influenced by environmental effects,
such as solvation or electron exchange with the surroundings.
Additionally, thermodynamic considerations are excluded; the focus
is strictly on the probabilistic theory of quantum mechanics.
Furthermore, the shape of the physical domain $\Omega$, which is
defined by its volume and the corresponding surface area, remains
fixed, along with the external potential generated by the nuclei
within the molecule.

The second section of this report analyzes the mathematical
framework and solutions to the problem. The third section includes
the discussion and concluding remarks.

\vskip 10mm

\noindent {\large {\bf II. THE STATISTICAL MODEL}}

\vskip 3mm

We will begin by introducing the concept of a physical domain within
a molecular system and the framework in which the model operates,
outlining its mathematical and physical aspects.

We define the domain mentioned above as a collection of atoms
organized within a chemical functional group, an arbitrary moiety,
or even as an individual atom within the molecular structure.  Each
domain represents a three-dimensional region within the molecular
framework and correlates to one of these units based on the specific
focus of the study. Since each domain is integrated into the overall
molecular structure, it operates as an open quantum system. The
transfer of charge between interacting domains results in a
non-integer number of electrons within the molecule. Various
methodologies are used to describe the physical domains in a
molecule. Some of these approaches have semi-empirical foundations, 
$^{13,14}$ whereas the only method with rigorous
theoretical foundations is the Quantum Theory of Atoms in Molecules
(QTAIM). $^{7,8}$ Some examples of these
domains are, for instance, $CO$ and $OH$ groups in formaldehyde and
methanol, respectively; $^{15,16}$ the $V$ metal atom in
the $[V(CO)_{6}]^{-2}$ cluster; $^{17}$ or $CC$ 
moiety in ethylene. Therefore, each domain $\Omega$ contains one or
more atoms. The number of electrons in the isolated domain is denoted by $N$,  which is defined 
as $N=Z_{\Omega}=\sum_{k\subset \Omega} Z_{k}$. Here,  $Z_{\Omega}$ represents the sum of 
atomic numbers $Z_{k}$ of the atoms which constitute the domain. The electron population of 
the domain can vary within a range of the integer values  $Z_{\Omega} \pm q$ where the integer
number $q$ stands for a parameter whose physical meaning is the
maximum charge that the domain can either accept or donate.

The electronic population of a region $\Omega$, denoted as
$N_{\Omega}$, $^{15,16}$ is defined using a non-integer
value $\mathcal{N}=N+\nu$. This value reflects the acceptor/donor
characteristics of the domain, where $N$ represents the integer
part, and $\nu$ is a real number within the interval $-q \leq \nu
\leq q$. The variable $\nu$ corresponds to the fraction of charge
that has been transferred to or from the domain.

Energy convexity does not hold for systems with a fractional number 
of particles, unlike systems with a fixed integer number of electrons, 
as discussed previously. $^{1,2,4}$ This situation necessitates an
extension of the density matrix $D$, which represents the quantum
mixed state in a molecular domain. This extension must accommodate
more than two pure states, each corresponding to a different number
of particles $M$, as illustrated in the case of convexity.
Therefore, $D$ can be expressed as follows

\begin{equation}
D=\sum_{M}\sum_{k} \omega_k^M \ket{\Phi_{k}^M}\bra{\Phi_k^M}=\sum_M
\sum_k \omega_k^M \hspace{0.05cm}{}^{M}\hspace{-0.1cm} D_k,
\label{1}
\end{equation}

\noindent where ${}^{M}D_k=\ket{\Phi_k^M}\bra{\Phi_k^M}$ represents
the $k$-th $M$-electron pure state within an antisymmetric
$M$-electron Hilbert space $\mathcal{H}_M$, corresponding to the
Hamiltonian eigenstates. The statistical weights that describe the
electron distribution are denoted by $\omega_k^M$, which satisfy

\begin{equation}
\sum_M \sum_k \omega_k^M =1 \hspace{0.5cm} 0\leq \omega_k^M \leq 1.
\label{2}
\end{equation}

\vskip 3mm

The density matrix $D$ lives in the Fock Space $\mathcal{F}$, i.e.,
$D\in \mathcal{F}$, where $\mathcal{F}$ is built as a direct sum of
$M$-particle Hilbert spaces $\mathcal{H}_M$ $^{18}$

\[
\nonumber \mathcal{F}=\bigoplus_{M=0}^{\infty}\mathcal{H}_M
\]

\vskip 3mm

We now present a three-state model that simplifies and, to some
extent, generalizes previous treatments by incorporating the
parameters $\nu$ and $q$. $^{12}$ The statistical
weights $\omega^{M}_{k}$ are explicitly defined as functions of
these two parameters represented as $\omega^{M}(\nu/q)$, as derived
algebraically. $^{12}$ In this model, we focus solely
on electronic ground states. Therefore, we will replace the label
$k$, which is associated with the quantum number in the statistical
weights, with $0$.

The three-states model retains only three states, a central
(neutral) state that contains $N$ electrons along with two {\it
edge} states, which are either cationic or anionic with a maximum
$\pm q$ electrons. This results in total electron counts of $N+q$ or
$N-q$, respectively. As a result, we perform a convex expansion for
the ground states $^{M}D_{0}$ of $M$-particle systems. Hence, the
quantum state of the molecular domain $\Omega$ is expressed as
follows

\vskip 3mm

\begin{equation}
D= \omega^{N-q}_{o}(\nu/q)\; {^{N-q}D}_{0} +\;
\omega^{N}_{o}(\nu/q)\;{^{N}D}_{0} +\; \omega^{N+q}_{o}(\nu/q)\;
{^{N+q}D}_{0}. \label{3}
\end{equation}

\vskip 3mm

\noindent A previous report provided solutions to the equations for
the zeroth and first-order moments of the electron distribution,
derived from an algebraic perspective. The zero- and first-order
moments stand for the normalization condition and the occupation
(electronic population) of the domain $\Omega$, $N_{\Omega}$, and
establishes that the mean value of the number of particles
$\left\langle M \right\rangle$, respectively. They define algebraic
relations for the statistical weights $^{12}$ by

\vskip 5mm

\begin{subequations}
\begin{equation}
\hspace{-2.5cm}\omega^{{N-q}}_{o} +\; \omega^{{N}}_{o} +\; \omega^{{N+q}}_{o}\; = 1
\hspace{1.0cm} normalization\; condition
\end{equation}
\begin{equation}
\omega^{{N+q}}_{o}\; -\; \omega^{{N-q}}_{o}\; =\;  \frac{\nu}{q} \hspace{1.0cm}
average\; number\; of\; particles\; condition
\end{equation}
\label{4}
\end{subequations}

\vskip 5mm

\noindent which provides solutions for the statistical $\omega^{N\pm
q}_{o}$ in terms of the central weight $\omega^N_{o}$ and the ratio
$\nu/q$, $^{12}$ which reads

\begin{equation}
\omega^{{N \pm q}}_{o}(\nu/q)=\frac{1 \pm \nu/q - \omega^{N}_{o}(\nu/q)}{2},
\label{5}
\end{equation}

\vskip 3mm

The approach presented here aims to determine the coefficients of
the $D$ expansion using the maximum entropy principle. As a result, 
Consequently, the output states considered are not all accessible, unlike 
those derived from the algebraic method; they are instead limited to the 
equilibrium ground states, which constitute a subset of all possible states. This 
complementary methodology
offers an advancement through the direct relationship established
between the transfer charge ratio and the chemical potential, which
will be derived in what follows.

The procedure aims to maximize the entropy $S$

\begin{equation}
S=\; - \sum_{M=N,N \pm q}\omega^{{M}}_{o}\; ln(\omega^{{M}}_{o})
\label{6}
\end{equation}

\vskip 5mm

\noindent given the available information for this open system,
known as the domain. The constraints include the normalization
condition $\sum_{M=N,N \pm q}\omega^{{M}}_{o}= 1$ and the average
number of particles (population) in the domain, expressed as
$\sum_{M=N,N \pm q}\omega^{{M}}_{o} M=\; \mathcal{N}=\; N+\nu$. We
refer to this group as the $\mathcal{N}$-{\it ensemble}. The
expression for maximization is commonly stated as follows

\begin{equation}
\delta \left [S - \gamma  \left(\sum_{M=N,N \pm q}\omega^{{M}}_{o} M -\; \mathcal{N} \right) -
\psi \left(\sum_{M=N,N \pm q}\omega^{{M}}_{o} - 1 \right) \right] = 0
\label{7}
\end{equation}

\vskip 5mm

\noindent with $\gamma$ and $\psi$ the associated Lagrange
multipliers. Therefore, the solutions are

\vskip 5mm

\begin{subequations}
\begin{equation}
\omega^{{N}}_{o} = \left[ 1+2\; cosh(q\gamma)\right]^{-1}
\end{equation}
\vskip 2mm
\begin{equation}
\omega^{{N\pm q}}_{o} =\; exp(\mp q\gamma)\; \omega^{{N}}_{o}
\end{equation}
\label{8}
\end{subequations}

\vskip 5mm

\noindent By substituting these results into Eq. (4b), we can
determine the net charge within the domain.

\begin{equation}
\nu = -2\; q\; \frac{sinh(\gamma q)}{ 1+2\; cosh(\gamma q)}
\label{9}
\end{equation}

\vskip 10mm

\noindent {\large {\bf III. DISCUSSION AND FINAL REMARKS}}

\vskip 3mm

The justification for the three-state model is based more on physical reasoning than on mathematical principles, 
supported by our current understanding of the system. This understanding is conveyed through equations (4a) and (4b), 
the normalization condition (which reflects the conservation of probability), and the electron population within the 
domain. The maximum charge occupation, denoted as $\pm q$, allowed by a domain represents a symmetrical framework that 
corresponds to its response to various molecular environments. This reveals the domain's consistently acceptor or donor 
character. Introducing additional states beyond the three already considered would necessitate adding them in pairs, 
resulting in an expansion of the Fock space dimension. While this is mathematically feasible, it would require knowledge 
of higher moments of the distribution to establish the statistical weights, which are not easily determined a priori. 

The variations of the coefficients in the $D$ expansion and the net
charge in the domain $\nu$ as a function of the physical magnitude
$\gamma$ are shown in Figures 1 and 2, respectively. They describe
the charge transference among the domains in the whole system.

The key analysis of these graphics lies in recognizing the
equivalence between the current statistical formulation and the
algebraic framework derived from Eqs. (4) that were previously
reported. $^{12}$ The current approach focuses
exclusively on equilibrium ground states. In this context, the
corresponding states within the aforementioned algebraic formalism
are the pure states denoted as ${^{N \pm q}D}_{0}$, as well as the
ground states represented by mixed states, which are illustrated as
straight lines characterized by coefficients  $\nu$ and $1 \mp \nu$.
These lines define the ground states for systems with a fractional
number of electrons, such as an $\Omega$-domain (see Eq. (4) of
Refs. $^{1,4}$

The coefficients of the ionic configurations $\omega^{{N \pm
q}}_{o}$ in the expansion $D$ (as shown in Eq. (3)) represent the
statistical measure of such configurations in the domain state. Therefore, these
coefficients can be interpreted as indicators of the acceptor or
donor character of the domains, depending on their values. As
$\gamma$ approaches either negative or positive infinity  $\gamma
\rightarrow \mp \infty$,  the coefficients $\omega^{N \pm q}_{o}$
tend to 1. In this scenario, ${^{N \pm q}D_{0}}$ becomes the
dominant state in the expansion, effectively representing the
limiting states of acceptor or donor behavior for the domain,
respectively.

As a result, it is important to note that the coefficient $\omega^{N}_{o}$ is 
never reaches unity, indicating that there are no $N$-particle pure states within 
this statistical description. Additionally, when 
$\gamma \pm \infty$, the system approaches the boundaries of each donor or acceptor 
characteristic $\mp q$ within the domain, meaning $\nu \rightarrow \mp q$. These limiting 
cases may arise in scenarios  that represent impossible accessible physical situations. When 
$\gamma \approx 0$, it indicates that the domain neither accepts nor donates charge. In this 
case, all coefficients have the same value of $\frac{1}{3}$, which defines a uniform statistical 
state. Therefore, $\gamma$ appears to influence the system's ability to donate or accept, 
allowing it to release or attract electrons.

Upon examining the variation of $\nu$ in Figure 2, the concept of
electronegativity $\chi$ becomes clear, which refers to the ability
to attract electrons. As $\gamma$ decreases (i.e., becomes more
negative), the system behaves more like an electron acceptor,
meaning it attracts more electrons when $\nu > 0$ and $\gamma$ is
increasingly negative. As the parameter $\gamma$ increases
positively, the domain transforms into a donor  characterized by
$\nu < 0$. Namely, negative values of $\gamma$ indicate an
attractive interaction of the domain, where positive values suggest
a tendency to release electrons. This behavior is similar to the
phenomenological concept of the {\it chemical potential} $\gamma$ as
it relates to a specific domain within the entire molecular system.
Traditionally, it is approximately equated to {\it the negative of the
electronegativity} represented as $\chi=-\gamma$. $^{19}$ In this 
model, regarding the behaviour discussed, it is a natural choice. 

The final topic for discussion is the equivalence between two methodologies: the previously 
established algebraic method $^{12}$ and the current statistical approach. This 
equivalence can be demonstrated by substituting Eqs. (8) and (9) into the algebraic equation (5) 
for the coefficients $\omega^{{N\pm q}}_{o}$. After this substitution, the coefficients align 
with those obtained from the current model. Thus, the present 3-state statistical model is 
equivalent to the previously defined 3-state algebraic model, as outlined in Eq. (9).  
$^{12}$

\newpage

\noindent \noindent {\large {\bf ACKNOWLEDGMENTS}}

\noindent This report has received financial support from Project
PIP No. 11220090100061 (Consejo Nacional de Investigaciones
Cient\'{\i}ficas y T\'ecnicas, Rep\'ublica Argentina). The author
thanks the Department of Physics, Facultad de Ciencias Exactas y
Naturales, Universidad de Buenos Aires for providing facilities.

\newpage

\noindent {\large {\bf REFERENCES}}

\vskip 5mm

\noindent $^{1}$ J. P. Perdew, R. G. Parr, M. Levy, J. Balduz, Jr., Phys. Rev. Lett. {\bf 49}, 1691 (1982).

\noindent $^{2}$ R. G. Parr and W. Yang, {\it Density-Functional Theory of Atoms and Molecules} 
(Oxford Univesity Press, New York, 1989).

\noindent $^{3}$ P. Geerlings, F. De Proft, W. Langenaeker, Chem. Rev. {\bf 103}, 1793 (2003) and references. 

\noindent $^{4}$ R. C. Bochicchio, D. Rial, J. Chem. Phys. {\bf 137}, 226101 (2012) and references.

\noindent $^{5}$ N. H. Cohen, A. Wasserman, J. Phys. Chem. A {\bf 111}, 2229 (2007) and references. 

\noindent $^{6}$  R. C. Bochicchio, Theor. Chem. Acc. {\bf 134}, 138 (2015) and references.

\noindent $^{7}$ R. F. W. Bader, {\it Atoms in Molecules: A Quantum Theory} ( Oxford University Press, Oxford, 1994).

\noindent $^{8}$ P. Popelier,  {\it Atoms in Molecules: An Introduction} (Pearson Edu, London, 1999). 

\noindent $^{9}$ A. Mart\'{\i}n Pend\'as, E. Francisco, J. Chem. Theory Comp. {\bf 15}, 1079 (2019).

\noindent $^{10}$ G. Acke, S. De Baerdemacker, A. Mart\'{\i}n Pend\'as, P. Bultinck, WIREs Comput. Mol. Sci. {\bf 10}, 
e1456 (2019).

\noindent $^{11}$ K. Blum, {\it Density Matrix Theory and Applications} (Plenum Press, New York, 1981).

\noindent $^{12}$ R. C. Bochicchio, B. Maul\'en, J. Chem. Phys. {\bf 159}, 234111 (2023). 

\noindent $^{13}$ D. R. Alcoba, L. Lain, A. Torre, R. C. Bochicchio, J. Comp. Chem., {\bf 27}, 596 (2006) and references

\noindent $^{14}$  Y. Grin, A. Savin, B. Silvi, in {\it The Chemical Bond: Fundamental Aspects of Chemical Bonding}
Chap. 10, G. Frenking and S. Shaik (Eds.) (Wiley‐VCH Verlag, 2014)

\noindent $^{15}$ R.C. Bochicchio, L. Lain, A. Torre, Chem. Phys. Lett. {\bf 374}, 567 (2003). 

\noindent $^{16}$ R. Bochicchio, L. Lain, A. Torre, Chem. Phys. Lett. {\bf 375}, 45 (2003). 

\noindent $^{17}$ R. C. Bochicchio, R. M. Lobayan, C. P\'erez del Valle, Int. J. Quant. Chem., {\bf 117}, e25876 (2018). 

\noindent $^{18}$ G. G. Emch GG (1972) {\it Algebraic Methods in Statistical Mechanics and Quantum Field Theory} 
(Wiley-Interscience, New York, 1972). 

\noindent $^{19}$  R. G. Parr, R. A. Donnelly, M. Levy, W. E. J. Palke, Chem. Phys. {\bf 68}, 3801 (1978).

\newpage

%%%%%%%%%%%%%%%%%%%%%%%%%%%%%%%%%%%%%%%%%%%%%%%%%%%%%%%%%%%%%%%%%%%%%%%%%%%

\begin{center}
\includegraphics[width=10cm]{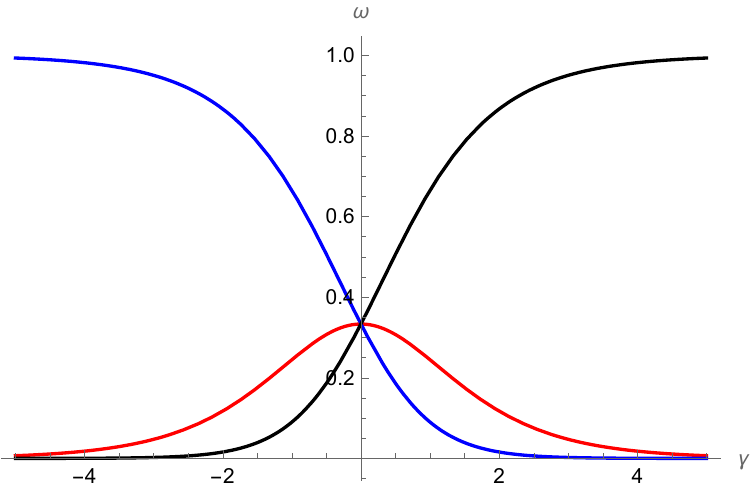}\\
\vskip 5mm
%\caption{Coefficients} \label{figura1}
Figure 1. Statistical coefficients $\omega^{M}_{o}$ as a function of $\gamma$.
$M=N$, red;  $M = N \pm q$, blue, black, respectively.
\end{center}

\newpage

\begin{center}
\includegraphics[width=10cm]{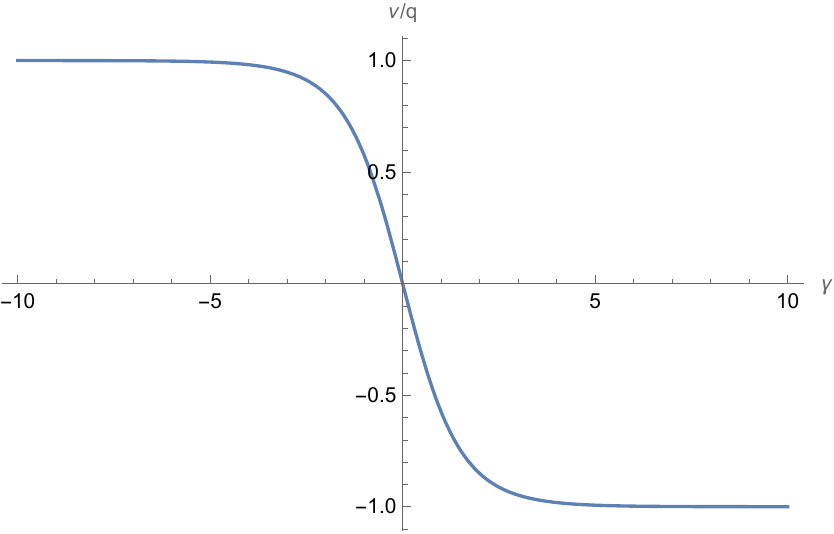}\\
\vskip 5mm
%\captionof{Transferred charge} \label{figura2}
Figure 2. Transferred charge $\nu/q$ as a function of $\gamma$.
\end{center}

\end{document}